\documentclass[showpacs,preprintnumbers,pre]{revtex4}
\usepackage{graphicx}
\usepackage{dcolumn}
\usepackage{bm}
\begin{document}
\title{
Field-driven solid-on-solid interfaces
moving under a stochastic Arrhenius dynamic: effects of 
the barrier height
}
\author{
G.~M.~Buend{\'{\i}}a$^a$}
\author{
P.~A.~ Rikvold$^{b,c}$}
\author{M.~Kolesik$^{d,e}$}
\affiliation{
$^a$Department of Physics, Universidad Sim{\'o}n Bol{\'{\i}}var, Caracas 1080, Venezuela\\
$^b$School of Computational Science, 
Center for Materials Research and Technology
and Department of Physics,
Florida State University, Tallahassee, Florida 32306-4350, USA \\
$^c$ National High Magnetic Field Laboratory, Tallahassee, Florida 32310, USA\\
$^d$Institute of Physics, Slovak Academy of Sciences,
Bratislava, Slovak Republic\\
$^e$Optical Sciences Center, University of Arizona,
Tucson, Arizona 85721, USA\\
}
\date{\today}
\begin{abstract}
We present analytical results and kinetic Monte Carlo simulations for the mobility and microscopic
structure of solid-on-solid (SOS) interfaces driven far from equilibrium by an external force, such as an 
applied field or (electro)chemical potential difference. 
The interfaces evolve under a specific stochastic dynamic with a local energy barrier (an Arrhenius dynamic), 
known as the transition dynamics approximation (TDA).
We calculate the average height of steps on the interface, 
the average interface velocity, and the skewness of the interface as functions of the 
driving force and the height of the energy barrier. We find that the microscopic interface structure depends quite 
strongly on the barrier height. As the barrier becomes higher, the local interface width decreases and the 
skewness increases, suggesting increasing short-range correlations between the step heights.    
\end{abstract}

\pacs{ 
68.35.Ct 
75.60.Jk 
68.43.Hn 
05.10.Ln 
}

\maketitle
\section{Introduction}
\label{sec:INTRO}

A large fraction of the processes essential to some of the most important technologies in our society 
involve the physical and chemical properties of interfaces: 
an interface is the place where molecules from different phases 
come into contact, and where chemical bonds are formed. 
Catalytic reactions and semiconductor devices are only two examples of systems where the role of surfaces and
interfaces is fundamental \cite{DUKE02,ERTL97}. 
It is therefore of the utmost importance to understand the fundamental processes occurring
at surfaces and interfaces in order to design more efficient mechanisms to contribute to the technological progress.

Since the specific dynamics of the evolution of surfaces and interfaces are in general not known,
it is useful to construct stochastic models that reproduce the essential features of the physical 
system. Until recently it was commonly believed that dynamics that respect the same conservation laws and all 
obey detailed balance, give essentially the same qualitative behavior. However, recent results clearly show 
that there are nonconservative dynamics that, although they all obey detailed balance and the same conservation laws, 
lead to very different interface microstructures \cite{RIKV00B,RIKV02,KATL04,ERDE04,BUEN05}. 
Since many interface properties, such as mobility and chemical activity, are determined by the 
microstructure, great care must therefore be taken in selecting the appropriate 
dynamic for the physical or chemical system of interest.

In this work we present analytical results and kinetic Monte Carlo (MC) simulations for the mobility and
microscopic structure of (1+1)-dimensional solid-on-solid (SOS) interfaces
\cite{BURT51} in an Ising lattice-gas model, which are 
moving far from equilibrium under an external force, such as an applied magnetic or electric field
or an (electro)chemical potential difference. The system evolves under an Arrhenius dynamic
which includes a microscopic barrier that represents a transition state inserted between the states 
allowed in the Hamiltonian. Arrhenius dynamics are appropriate when discrete Ising or lattice-gas 
models are used to simulate dynamics in an underlying, continuous potential. In such cases the transition state in
the Arrhenius dynamic could represent a saddle point in 
a corrugation potential for particle diffusion, or a high-energy state associated with a transitional 
spin state \cite{MITC02,ALAN92}. The dynamic considered in this paper is the two-step transition 
dynamic approximation (TDA) \cite{ALAN92,BUEN04}.
The transition-state energy is approximated by \cite{ALAN92}
\begin{equation}
E_T=\frac{E_f+E_i}{2} + \Delta \;,
\end{equation}
where $E_f$ and $E_i$ are the initial and final energies, and $\Delta$ is the microscopic energy
barrier. In electrochemical applications, this corresponds to the symmetric Butler-Volmer approximation 
\cite{SCHM96}. The transition rates for the TDA are given by
\begin{equation}
W_{\rm {TDA}}=\frac{1}{1+\exp[\beta(E_T-E_i)]}\frac{1}{1+\exp[\beta(E_f-E_T)]} \;.
\label{eq:wtda}
\end{equation}
The TDA belongs to the class of dynamics defined as ``hard," in which the transition rates cannot
be factorized into one part that depends only on the interaction energy and another that
depends only on the field energy. Dynamics that {\it have\/} 
this factorization property are defined as ``soft" \cite{MARR99}. 
When applied to kinetic Ising lattice-gas models, soft and hard Arrhenius dynamics give very different nucleation rates 
\cite{BUEN04}. It has also recently been shown that SOS interfaces evolving under soft Arrhenius dynamics have quite 
different structures that those evolving under hard Arrhenius dynamics \cite{BUEN05}. 
In soft Arrhenius dynamics, the barrier height only results in a change in the overall time scale of the
simulation and has no effect on the interface microstructure \cite{BUEN05}. 
In the present work we
focus on the effect of the barrier height $\Delta$ on the mobility and microstructure of interfaces moving under the
hard TDA dynamic.

\section{The Model}
In this section we give a brief description of the SOS model and the analytical theory for the interface
microstructure and mobility. Details can be found in Refs.~\cite{RIKV00B,RIKV02,BUEN05,BURT51}. 

The SOS interfaces are described by the nearest-neighbor Ising Hamiltonian,
\begin{equation}
{\cal H} = -\sum_{x,y} s_{x,y} \left( J_x s_{x+1,y} + J_y s_{x,y+1} 
+ H \right) 
\;, 
\label{eq:ham}
\end{equation}
where $s_{x,y}=\pm1$, $\sum_{x,y}$ runs over all sites, and 
 the applied field $H$ is the driving force. The interface is introduced by fixing
$s_{x,y}=+1$ and $-$1 for large negative and positive $y$, respectively.  
We take $H \ge 0$, such 
that the interface on average moves in the positive $y$ direction. 
This Ising model is equivalent to a lattice-gas model with local
occupation variables  $c_{x,y} \in \{0,1\}$ (for the exact relations between the parameters in the two models, see,
e.g., Ref.~\cite{BUEN05}). 

The interface is described by a single-valued integer function $h(x)$  
with steps $\delta(x)=h(x+1/2)-h(x-1/2)$ at integer values of $x$. An example of an SOS
interface is shown in Fig.~1. The spins in the anisotropic square-lattice Ising model can
be divided into classes, labeled $jks$, where $j$ and $k$ are the number of broken bonds
between the spin and its nearest neighbors in the $x$ and $y$ directions, respectively, 
and $s$ is the spin value. Only spins that belong to the classes 01$\pm$, 11$\pm$, and 21$\pm$ can flip in
the SOS interface. Whenever a spin flips from $-1$ to $+1$, the corresponding column of the
interface advances by one lattice constant in the $y$ direction. 
Conversely, the column recedes by one lattice constant when a spin flips from $-1$ to $+1$. 
In this approximation the spin-class populations on both
sides of the interface are equal, and the contribution to the mean
velocity in the $y$ direction from sites in the classes $jk-$ and $jk+$ becomes
\begin{equation}
\langle v_y(jk) \rangle 
= 
W \left( \beta \delta E(jk-) ,\beta \Delta  \right)
-
W \left( \beta \delta E(jk+) , \beta \Delta \right) 
 \;. 
\label{eq:generalv}
\end{equation}
where $\delta E(jk\pm)$ is the change in energy due to flipping a spin in the class
$jk\pm$, and the transition rates $W$ are calculated from Eq.~(2).
The mean propagation velocity perpendicular to the interface is defined as  
\begin{equation}
\langle v_\perp (T,H,\phi) \rangle 
= 
\cos \phi \sum_{j,k} \langle n(jks) \rangle \langle v_y (jk) \rangle 
\;, 
\label{eq:totalv}
\end{equation}
where $\phi$ is the overall angle between the interface and the $x$ axis. 
In this mean-field approximation, the heights of the individual steps are assumed to be statistically independent. 
As a consequence, the mean spin-class populations $\langle n(jks) \rangle$ are obtained from the product of the
step-height probability density function (pdf) for $\delta(x)$ and $\delta(x+1)$. 
The pdf is given by the interaction energy corresponding to 
the $|\delta(x)|$ broken $J_x$ bonds between spins in the columns centered at $(x-1/2)$ and $(x+1/2)$ as
\begin{equation}
 p[\delta(x)] = Z(\phi)^{-1} X^{|\delta(x)|}
\ e^{ \gamma(\phi) \delta(x) } \;. 
\label{eq:step_pdf}
\end{equation}
The factor $X$ determines the width of the pdf; in equilibrium it is simply the Boltzmann
factor, $e^{- 2 \beta J_x}$. The quantity
$\gamma(\phi)$ is a Lagrange multiplier that maintains the mean step 
height at an $x$-independent value, $\langle \delta(x) \rangle = \tan \phi$.
The partition function is 
\begin{equation}
Z(\phi)
=
\sum_{\delta = -\infty}^{+\infty} X^{|\delta|} e^{ \gamma(\phi) \delta } 
=
\frac{1-X^2}{1 - 2 X \cosh \gamma(\phi) + X^2} \;,
\label{eq:Z}
\end{equation}
where $\gamma(\phi)$ is given by 
\begin{equation}
e^{\gamma (\phi)}
= 
\frac{ \left(1+X^2 \right)\tan \phi 
+ \left[ \left( 1 - X^2 \right)^2 \tan^2 \phi + 4 X^2 \right]^{1/2}}
{2 X \left( 1 + \tan \phi \right)} 
\;.
\label{eq:chgam}
\end{equation}
A non-linear response approximation based on a dynamic mean-field approximation for the equation of motion of the 
single-step pdf, together with a detailed-balance argument for the stationary state, gives the following expression 
for the field-dependent $X(T,H)$ \cite{RIKV00B,RIKV02},
\begin{equation}
X(T,H) = e^{-2 \beta J_x} 
\left\{
\frac{e^{-2 \beta H}W(21-) + e^{2\beta H}W(21+)}
{W(21-) + W(21+)}
\right\}^{1/2}
\;,
\label{eq:XTH}
\end{equation}
where the $W(jks)$ are the single-site transition rates associated with
the flipping of a spin of the class $jks$. For the details of the calculation see Refs.~\cite{RIKV00B,RIKV02}. 
For the TDA dynamic defined in Eq.~(2) one gets \cite{BUEN05}
\begin{equation}
X_{\rm TDA}(T,H) = 
e^{-2 \beta J_x} 
\left\{
\frac{e^{2 \beta J_x} \cosh(2 \beta H) + e^{-2\beta J_x} + 2 \cosh(\beta\Delta) \cosh(\beta H)}
{e^{-2 \beta J_x} \cosh(2 \beta H) + e^{2\beta J_x}+ 2 \cosh(\beta\Delta) \cosh(\beta H)}
\right\}^{1/2}
\;.
\label{eq:XTDA}
\end{equation}

\section{Results}
The numerical simulations were performed with a continuous-time $n$-fold way rejection-free algorithm 
\cite{BORT75} at $T=0.2T_c$ (where $T_c\approx2.269J$ is the critical
temperature for the isotropic, square-lattice Ising model) with $L_x=10\,000$ and isotropic interactions, $J_x=J_y=J$. 
In Fig.~2 we show the results for 
the average stationary step height $\langle |\delta| \rangle$ vs $H$, for different barrier values. The agreement 
between the analytical results and the MC data is very good for the smaller values of $\Delta$. We believe that 
the differences between the analytical and the MC results for 
the higher values of $\Delta$, particularly for small $H$, may be related to
the fact that the assumption of independent step heights is not valid for larger barriers due 
to the increasing presence of short-range correlations. 
Figure~2 indicates that the local width of the interface depends strongly on 
$H$ and on the height of the
barrier. The larger the barrier, the smaller the local interface width. 
We also note that the agreement is not very good for $H <
\Delta$. However, the agreement is almost exact for $H = \Delta$,
and it remains quite reasonable for $H>\Delta$. 

The dependence of the perpendicular interface velocity on the 
field is shown in Fig.~3 for several
values of $\Delta$. The agreement between the analytical and MC 
results for the velocity appears to be very good for all values of
$\Delta$ and $H$. 
This is, however, somewhat of an optical illusion. The good  
agreement occurs for the higher velocities, which all are attained for 
$H \ge \Delta$. If one concentrates on the low velocities attained
for $H < \Delta$, one indeed finds that the theory significantly
underestimates the simulation data. This is as expected, since the
dominant contribution to the velocity at this low temperature is
from spins in class $11s$, whose abundance in the simulated
interface is larger than expected from the theory.
As predicted by the theory, the interface velocity of systems 
that evolve under the TDA is bounded by unity, contrary to 
the case of the soft Arrhenius dynamic considered in 
Ref.~\cite{BUEN05}, for which it increases exponentially with $H$. 
From the figure it is clear that, as the barrier
increases, the velocity increases more slowly with $H$, 
and it reaches its maximum value at higher fields.

As mentioned above, the 
analytical predictions for the class populations are based on the mean-field assumption that the steps are 
statistically independent. Within this approximation, the average populations of each class of spins must be 
the same in front of 
and behind the interface. However, the simulation results clearly show that this is not the case in general. As an example, the six mean class populations -$\langle n(01s) \rangle,  \langle n(11s) \rangle$, and $\langle n(21s) \rangle $ with $s=\pm1$ - are shown vs $H$ in Fig.~4, for $\Delta=2J$. Again we notice that, for this relative high value of $\Delta$, there is not a complete 
agreement between the theoretical and the MC data at intermediate fields.
The results indicate the presence of short-range correlations between neighboring steps that cause differences between the spin 
populations on the leading and trailing edges of the interface.
When $\langle n(21-) \rangle > \langle n(21+) \rangle$, the interface is characterized by a broadening of protrusions 
on the leading edge (``hilltops"). The relative skewness can be quantified by the following quantity \cite{NEER97}
\begin{equation}
\rho = \frac{\langle n(21-) \rangle - \langle n(21+) \rangle}
{\langle n(21-) \rangle + \langle n(21+) \rangle}
\;.
\label{eq:rho}
\end{equation}
This skewness parameter is shown in Fig.~5 for different values of the barrier 
height. Notice that the relative skewness shows a dramatic increase as the
barrier height becomes larger, particularly for low and intermediate fields. These
results suggest that the assumption of independent step heights fails for systems with high barriers. 
Skewness has also been observed in other SOS-type models \cite{NEER97,PIER99,KORN00C}.

\section{Conclusions}
In this work we have presented numerical and analytical results for the mobility and interface microstructure of an 
SOS interface, which is driven far from equilibrium by an applied field. The system evolves according to an Arrhenius 
dynamic known as the transition dynamics approximation (TDA). The Arrhenius dynamic interposes between the 
Ising lattice-gas states a transition state representing a local energy barrier. In this study we focus on the effects 
of the barrier height on the interface microstructure and mobility.
The analytical results are obtained with a non-linear response theory, which assumes that there are no correlations 
between the heights of different steps on the interface. 

We found that the microscopic properties of the interface depend quite strongly on the barrier height. The interfaces 
are less wide, move more slowly, and are  more skewed as the barrier height increases.
The agreement between theory and simulation is very good for the
higher interface velocities (Fig.~3). 
For low barriers, the local interface 
width is also in good agreement, but for larger $\Delta$, the
agreement only becomes good for  
$H \ge \Delta$ (Fig.~2). For higher barriers, 
the results suggest that short-range correlations become important 
(Fig.~4,~5). 
To account for such correlation effects, 
an improved theory would be needed. 
Finally we emphasize that the disagreements between theory and simulations
presented in this paper represent a ``worst-case scenario." The agreement
improves considerably for higher temperatures (below $T_c$), and 
at $T=0.6T_C$ it is quite good for all barrier heights and fields studied.

\section*{Acknowledgments}
\label{sec:ACK}
G.~M.~B.\ gratefully acknowledges the hospitality of the 
School of Computational Science at Florida State University. 
This research was supported in part 
by U.S.\ National Science Foundation Grant Nos.~DMR-0240078 and DMR-0444051,
by Florida State 
University through the Center for Materials Research and Technology and
the School of Computational Science, by the U.S.\ National High Magnetic Field Laboratory, and
by the Deanship of Research and Development of Universidad Sim{\'o}n Bol{\'{\i}}var.
                                                                                

\clearpage
                                                                                
\begin{figure}[ht] 
\includegraphics[angle=0,width=.45\textwidth]{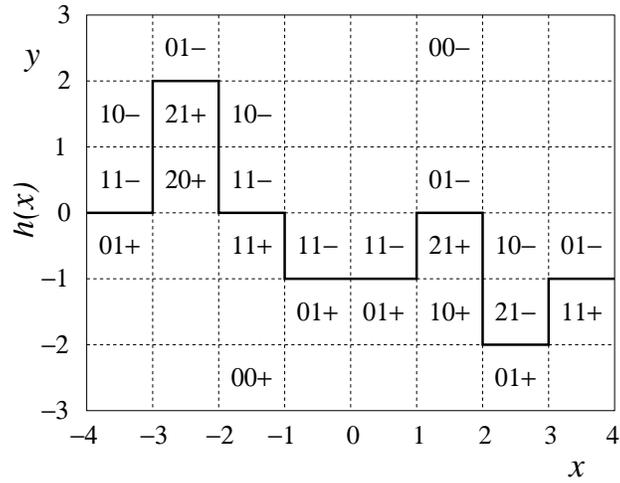}
\caption[]{
A segment of an SOS interface $y=h(x)$ between a positively
magnetized phase (or ``solid'' phase in the lattice-gas picture) 
below and a negative (or ``fluid'') phase
above. Sites in the uniform bulk phases are
$00-$ and $00+$. 
From Ref.~\protect\onlinecite{RIKV02}. 
}
\label{fig:pict}
\end{figure}


\begin{figure}[ht] 
\vspace{0.5truecm}
\includegraphics[angle=0, width=.45\textwidth]{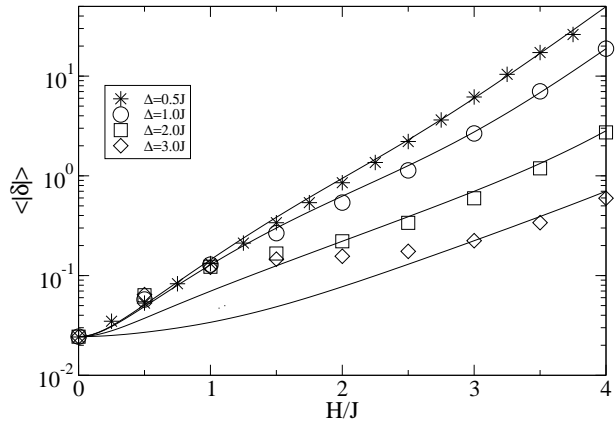}\\
\caption[]{
Average stationary step height 
$\langle | \delta | \rangle$ 
vs $H$ for $\phi$=0 at $T = 0.2T_c$.
The curves represent the analytical results. 
The MC data were obtained directly by summation 
over the simulated single-step pdfs.  
}
\label{fig:dh}
\end{figure}


\begin{figure}[ht]
\includegraphics[width=.45\textwidth,angle=0]{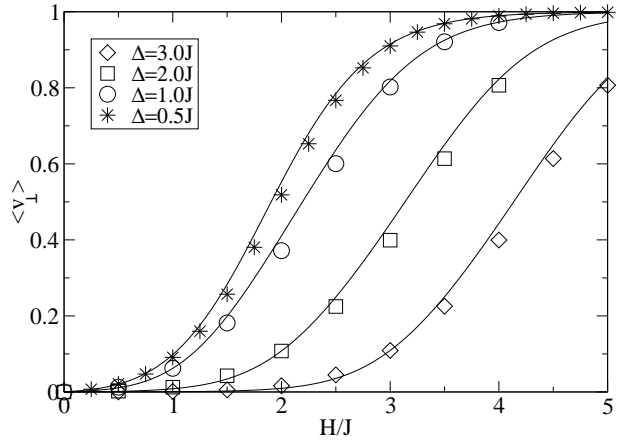}\\
\caption[]{
The average stationary
normal interface velocity $\langle v_\perp \rangle$ vs $H$ at $T = 0.2T_c$, calculated 
 for several values of the barrier $\Delta$. The symbols represent MC data, and 
the solid curves analytical results.
}
\end{figure}

\begin{figure}[ht]
\vspace{0.5truecm}
\includegraphics[width=.45\textwidth,angle=0]{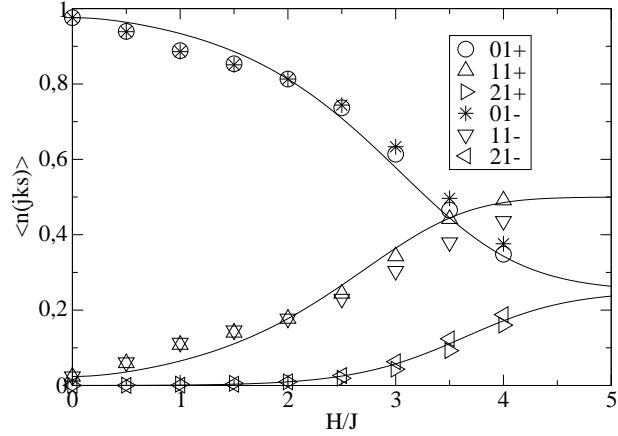}\\
\caption[]{
Mean stationary
class populations $\langle n(jks) \rangle$ vs $H/J$ at $T = 0.2T_c$, for
$\Delta=2J$. The symbols represent MC data, and
the solid curves analytical results.
}
\end{figure}

\begin{figure}[ht]
\vspace{0.5truecm}
\includegraphics[angle=0, width=.45\textwidth]{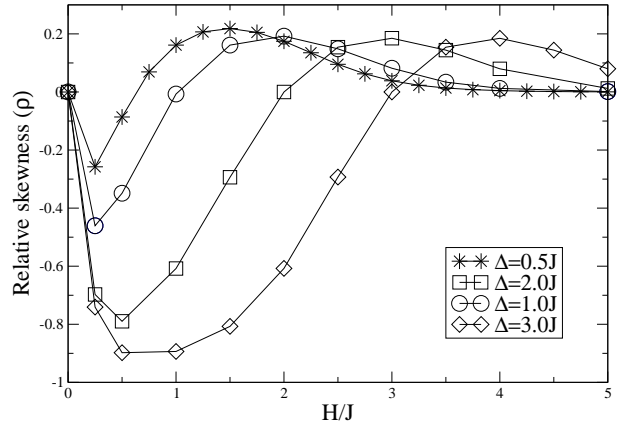}\\
\caption[]{MC data for the relative skewness $\rho$, defined in Eq.~(6), vs $H$ at $T = 0.2T_c$, for different
values of the barrier height $\Delta$.
The solid lines are merely guides to the eye. 
}
\label{fig:rho}
\end{figure}

\end{document}